# Sex-specific ultraviolet radiation tolerance across *Drosophila*

James E. Titus-McQuillan[1*], Brandon A. Turner[1], Rebekah L. Rogers[1]

[1]Department of Bioinformatics and Genomics, University of North Carolina, Charlotte NC, USA.
[*]Corresponding author: [1]Department of Bioinformatics and Genomics, University of North Carolina, Charlotte NC, titusmcquillan@gmail.com

**Abstract:**

The genetic basis of phenotypic differences between species is among the most longstanding questions in evolutionary biology. How new genes form and the processes selection acts to produce differences across species are fundamental to understand how species persist and evolve in an ever-changing environment. Adaptation and genetic innovation arise in the genome by a variety of sources. Functional genomics requires both intrinsic genetic discoveries, as well as empirical testing to observe adaptation between lineages. Here we explore two species
of *Drosophila* on the island of São Tomé and mainland Africa, *D. santomea* and *D. yakuba*. These two species both inhabit the island, but occupy differing species distributions based on elevation, with *D. yakuba* also having populations on mainland Africa. Intrinsic evidence shows genes between species may have a role in adaptation to higher UV tolerance with DNA repair mechanisms (*PARP*) and resistance to humeral stress lethal effects (*Victoria*). We conducted empirical assays between island *D. santomea*, *D. yakuba*, and mainland *D. yakuba*. Flies were shocked with UVB radiation (@ 302 nm) at 1650-1990 mW/cm$^2$ for 30 minutes on a transilluminator apparatus. Custom 5-wall acrylic enclosures were constructed for viewing and containment of flies. All assays were filmed. Island groups did show significant differences between fall-time under UV stress and recovery time post-UV stress test between regions and sex. This study shows evidence that mainland flies are less resistant to UV radiation than their island counterparts. Further work exploring the genetic basis for UV tolerance will be conducted from empirical assays. Understanding the mechanisms and processes that promote adaptation and testing extrinsic traits within the context of the genome is crucially important to understand evolutionary machinery.

## Introduction

The evolutionary response to shifting selective pressures is among the most profound open questions in evolutionary theory (Darwin, 1859; Dobzhansky, 1937). The way that organisms develop novel phenotypes that allow them to invade new habitats or survive under environmental shifts is essential to understand the outcomes and trajectory as organisms respond to new selective regimes. If adaptation can follow only a few paths to survival, then we expect species to show high rates of convergent evolution (Stern, 2013). However, if multiple different modes of phenotypic change allow survival in the face of environmental shifts, then rates of convergence may be lower (Emery & Clayton, 2004; Whittall et al. 2006; Zhou et al. 2008; Reviewed in Rosenblum et al. 2014). In such a case, the existence of multiple paths to success may allow more species to invade open niches more readily than if all must follow a single phenotypic (or genetic) solution (Kauffman & Levin, 1987; Schoville et al. 2012). Species that harbor greater amounts of standing genetic and phenotypic variation may be more adept at invading open niches than species with limited variation (Hermisson & Pennings, 2005). If standing variation is readily available, in theory, can offer an instant reservoir of phenotypic and genetic variation that is available when species experience selective shifts.

Island invasion allows us to study the evolutionary response to discrete shifts in habitat where we can observe the phenotypic and genotypic response to environmental changes (Reviewed in Brown et al. 2013). As species invade new habitats, alternative phenotypes are expected to form and spread, facilitating adaptive changes. To better understand how organisms generate novel traits in the face of environmental change, this study aims to empirically test if environmental factors have led to local adaptation in a non-model *Drosophila* from the island of São Tomé. *Drosophila santomea* is an endemic species from the island country of São Tomé and Príncipe, found at high elevation, not found below 1150 m on the volcano Pico de São Tomé [2024 m] (Lachaise et al. 2000). It has differentiated from its mainland African progenitor populations and all other species in the *melanogaster* group, in that it has little to no abdominal pigmentation, even along elevational gradients (Lachaise et al. 2000; Llopart et al. 2002; Matute & Harris, 2013).

This paler phenotype is curious, as it contradicts expectations for high altitude adaptation and other known solutions for survivorship under increased UV exposure. Darkening as a mechanism to combat harmful UV radiation is commonly observed in many metazoan groups – pocket gophers (Goldman, 1947), lizards (Reguera et al. 2014), grasshoppers (Harris et al. 2013), wasps (de Souza et al. 2020), and even zooplankton (Ulbing et al. 2019). Even in other species of *Drosophila,* high levels of melanin may be a trait that confers UV resistance at higher elevations (Pool & Aquadro, 2007; Zhao et al. 2015). Previous work empirically testing UV tolerance on latitudinal clines has shown that UV has a significant effect on survivorship in *D. melanogaster* embryos (Svetec et al. 2016). Increasing melanin production confers protection from UV radiation by acting as an absorbent filter preventing penetration, and subsequent DNA damage (Brenner & Hearing, 2008). This "altitude-induced melanism" is observed commonly across diverse taxa, yet not in *D. santomea*.

Another species of *Drosophila* resides on the island, *Drosophila yakuba,* forming the *D. yakuba – santomea* complex. *D. yakuba* is a more recent secondary colonization event to the island of São Tomé (Cariou et al. 2001; Coyne et al. 2002; Obbard et al. 2012). It is estimated that approximately 400,000 years ago these two species diverged (Llopart et al. 2002) with clear phenotypic differences. A significant diagnostic feature between the two species is that *D. yakuba* has heavily pigmented abdominal segments among both sexes, unlike *D. santomea*, which lacks melanistic pigmentation. While *D. santomea* ranges above an elevation of 1150 m, *D. yakuba* inhabits São Tomé's lower elevations, where it is not found above 1400 m. The species also ranges across central mainland Africa. These species have an intermediate region on the island where known introgression takes place (Lachaise et al. 2000). Between these species elevational ranges, a hybrid zone occurs between 1150 m – 1400 m.

In this island model, it is hypothesized that local adaptation has redefined species ranges given the elevational restrictions. Matute and Harris (2013) empirically tested temperature tolerances between the two species residing on the island, and that temperature leads to physiological changes that confer lower fitness and fertility. *D. santomea* is found in mist forests with lower temperature, while *D. yakuba* is found in disturbed, mesic, or secondary forest habitat (Lachaise et al. 2000; Cariou et al. 2001). These species show partial, but not complete reproductive incompatibilities that include pre and post-zygotic isolation (Lachaise et al. 2000; Cariou et al. 2001; Coyne et al. 2002; Llopart *et al.* 2002; Obbard et al. 2012; Matute et al. 2009; Matute & Harris, 2013).

*Evolution of Sex-specific Traits*

The *D. yakuba – D. santomea* species complex shows sex-specific divergence from the island to the mainland. Males are paler than their island counterparts, but still retain some abdominal pigmentation (Llopart et al. 2002). In contrast, females have lost pigmentation at stripes along abdominal segments and the pigmentation spot on the lower abdominal bands (Llopart et al. 2002). Additionally, there is a known size asymmetry for males and females in *D. yakuba* and *D. santomea* (Llopart et al. 2002). Temperature has also been shown to potentially be sex-specific (Llopart et al. 2005a, b), as *D. santomea* female fertility is markedly lower than males at higher temperatures (Matute et al. 2009). Intraspecific plasticity to temperature in male *D. santomea* has an effect in genital ventral branches (Peluffo et al. 2021).

Mating preferences persist between the sexes given that F1 hybrid offspring nearly exclusively come from *D. yakuba* males and *D. santomea* females. The onus being on female *D. santomea's* preference. Where male *D. yakuba* have no preference courting females from either species (Coyne et al. 2005). Pre and post-zygotic isolation has a role in mating preference (Moehring et al. 2006a, b; Peluffo et al. 2021). Sexual isolation is driven, in part, by courtship and song as a sex-specific trait between the sexes of non-conspecifics (Watson et al. 2007; Blyth et al. 2008; Cande et al. 2012). Studies also observe signals of the large-X effect within the *D. yakuba – D. santomea* complex resulting in speciation (Llopart, 2012). This complex confers Haldane's Rule of sterility (Haldane, 1922; Coyne, 1985), through prior studies showing the X chromosome has a disproportionately large effect on hybrid male sterility (Coyne et al. 2004; Moehring et al. 2006b).

Theory suggests that pleiotropic effects can result in antagonistic evolutionary pressures if traits have sex-specific impacts on selection (Darwin, 1871; Fisher, 1931; Lande, 1980; Parker, 1979). Because different sexes employ alternative reproductive strategies, both risk tolerance and phenotypic outcomes that may diverge between males and females (Trivers, 1972; Lande, 1980; Parker, 1979; reviewed in Bonduriansky & Chenoweth, 2009). In these scenarios, a trait that may be beneficial in one sex could be detrimental in the other leading to antagonistic pleiotropy (Lande, 1980; Bonduriansky & Chenoweth 2009; Rice, 1987; Rowe et al. 2018). Sexually dimorphic traits, however, can remove or reduce such constraints as females and males may adapt to their behavioral and reproductive environment independently (Trivers, 1972; Lande, 1980; Albert & Otto, 2005). This freedom from antagonistic sex-biased constraints may be even more important if evolutionary risks experience different payouts in males and females (Trivers, 1972).

Because evolutionary modes, reproductive strategies, and molecular or biological background differ in the sexes, it is all the more imperative to include sex as a biological variable in phenotypic analyses (Lee, 2018). To identify sex-specific effects of survival we assay UV tolerance separately for males and females. As we report, this use of sex as a biological variable offers greater precision in our phenotypic assays and allows us to identify significant sex-by-species interactions that would be missed in single sex or mixed-sex assays. These types of sex specific assays are essential to fully understand the modes and outcomes of evolutionary processes.

This study uses *D. santomea* as a model to empirically test modes of adaptation to UV stress that are expected to occur at high altitude in nature compared to its sister taxa of *D. yakuba* from locally adapted island environments and original mainland populations. To help connect phenotypes to genotypes, we also examine gene expression patterns to determine whether genetic changes are potentially linked to UV tolerance. We hope to discern the change in island environments, via UV resistance, for traits hypothesized to differentiate *D. yakuba* and *D. santomea* that might confer adaptive phenotypic variation between the species. Integrating ecology, genomic signal, and empirical hypothesis tests on phenotypic traits gives a holistic view on this evolutionary phenomenon.

**Methods:**

*Drosophila lines*

A total of 24 isofemale lines consisting of two species spanning the island São Tomé and mainland Africa. We used 9 *Drosophila santomea* lines and 15 *Drosophila yakuba* lines between São Tomé (five lines) and mainland *D. yakuba* (ten control lines) used in this study. Fly lines from the island of São Tomé were provided by the Matute lab. Our mainland *D. yakuba* control lines were provided by the National *Drosophila* Species Stock Center.

*Collection*

For each isofemale line we generated experimental animals by allowing groups of parentals (around 10-20) to mate and lay eggs in fresh vials, containing 4 ml of standard food, placed into an incubator at 22°C and 24°C for *D. santomea* and *D. yakuba*, respectively. The incubator was programed with a 12:12 light:dark cycle and 50% humidity. Emerging virgin offspring were anaesthetized with $CO_2$ in the morning and placed in new vials with standard media, filtered by sex for each line. They were then allowed to mature for three days prior to experiment without dry baker's yeast, to prevent flies potentially becoming stuck, under the same incubator conditions above.

*Enclosure Design and Construction*

To view the UV tolerance of flies for UV trials, we constructed custom enclosures made of clear acrylic. Acrylic, as a material, is well suited for these experiments as it has superlative clarity and is easily manipulated, unlike other materials such as glass. We cut four identical side panels and one top panel with an open bottom out of a 3 mm thick sheet on clear acrylic using a GlowForge™ laser cutter. Panels are fused to form the enclosures using Weld-On® 3 solvent cement. The interior volume of the enclosures is 26 mm wide by 26 mm long by 19 mm high. For full schematics see supplementary materials (enclosure_schematic.pdf [for laser cutting importation] and enclosure dimensions).

*Experimental Design*

UV exposure was generated on an Analytik-Jena 8-watt UVP 3UV Transilluminator LMS-20 [95-0417-01 (US)]. We set the light spectrum to 302nm (UVB). The transilluminator was given time to warm up before conducting the experiment for around 15 to 25 minutes as flies were gathered. Our transilluminator has internal cooling, though the platform can still heat up. To mitigate any form of heat shock, we monitored the platform with a temperature probe and cooled the platform with a fan. Given that *D. santomea* live at cooler temperatures than most *D. melanogaster* conspecifics (Lachaise et al. 2000). Limitations of past work on UV tolerance, a known concern is the collection of other variables has required separate assessment in addition to the experiment itself like heat and desiccation (Matute & Harris, 2013; Svetec et al. 2016).

To accomplish temperature regulation, we directed a fan on the glass transilluminator plate, providing directed airflow across the surface. Because glass is an insulator, heat does not easily penetrate, and providing ample airflow allows for heat to dissipate more rapidly from the apparatus. Temperature readings occurred in real-time using a calibrated digital temperature probe. UVB incidence was measured 10 cm from the transilluminator platform after warm-up phase, per manufacturer's recommendation, and taken at multiple areas with a Solarmeter® Model 5.7 Sensitive UVA+B Meter. Flies were taken from the incubators and anaesthetized. Groups of 5 flies, used as replicates, were placed in an open bottom enclosure made of clear acrylic. Open bottom enclosures are essential to avoid filtering UV through acrylic during exposures.

Video recording was conducted on a Sony Alpha 6400 - APS-C Interchangeable Lens 24.2-megapixel Camera. The camera was attached to a ball head mount on a C-stand positioned approximately 2 to 3 feet above the transilluminator. We recorded at 1080p resolution in MP4 format using a Tamron 17-70mm f/2.8 Di III-A VC RXD Lens for Sony E mount. Field of view (FOV) was set so that all enclosures were in view and *Drosophila* specimens were in focus. To mitigate shaking and preserve image quality, without touching the camera while recording, we used the Sony Imaging Edge Mobile application. This allowed for remote viewing on a mobile device and remote operation of camera functions without the need to touch or disrupt the recordings. This function reduces the need for active hands-on personnel. The end result is an assay that reduces man-hours for experiments, enhances safety by preventing researcher UV exposure, and avoids unexpected incidents that might disrupt the recording of assay screens.

We performed 24 trials with 5 replicates of 5 flies for females, and 24 trials with 5 replicates of 5 flies for males. Maximum number of replicates per trial was 6 for strains NY73 and Cascade 19.16 females and Thera2005 males. Flies were exposed to an average UV incidence of 1849 $\mu W/cm^2$. Trial recordings lasted for approximately 30 minutes of video with a total file size of ~3.0 GB of mp4 files. Out of all trials, 121 replicates for males and 122 replicates for females were scorable from video footage. Flies were collected and monitored post-UV activity and categorized by recovery status. Status is categorized by active, barely active, inactive, and dead. Active is defined by a fly specimen that is actively moving around and could be flying. This category of fly is not ostensively discernable from a fly not under any stress. A barely active fly is moving but highly lethargic. A fly will be slow, may fall while mobile, and, while active, is not fully mobile. Inactive flies are not mobile; however, bodies will have some source of life such as twitching appendages. Usually, an inactive fly will show signs of activity given external stimuli, *e.g.* tapping the vial will yield slight twitches and movement. These flies are not motile. Finally, the dead category is defined as a fly that is no longer showing is discernable activity, without movement and any activity even given an external stimulus. Post-UV activity was recorded using a dissecting scope at 30-minute, one-hour, two-hour, four-hour, and 24-hour time intervals after filming.

*Geographic Reconstruction of Solar Stress*

Solar radiation map of the island of São Tomé was constructed using WorldClim 2.1 historical climate data (Fick et al. 2017), taking solar radiation raster layers at 30 seconds resolution. All raster layers were clipped by a mask, a shape file of the administration zone of São Tomé and Príncipe. In total 12 layers (corresponding to each month) were clipped by the shape file. Geo TIFF files were then converted to ascii files and overlayed to display a holistic annual solar radiation (kJ $m^{-2}$ $day^{-1}$) of São Tomé displayed in figure 1, where lighter color signifies higher solar radiation values. All clipping functions and raster translations were conducted in QGIS π (v. 3.14.1-Pi). The range map of *D. santomea* and *D. yakuba* was adapted from Lachaise et al. 2000 using vector graphics in Adobe Photoshop CC and Illustrator CC.

*Statistical Analysis*

All statistical analyses were conducted in R v4.2.2 (R Core Team, 2022) using rStudio 2022.12.0 Build 353 (Rstudio Team, 2020). For our real-time fly activity under UV stress, statistical tests were conducted among and between factors sex and region and their interaction. We used a Wilcox test (Bauer, 1972; Hollander & Wolfe, 1973) independently on male and female flies to compare if groups of flies originating from different geographic regions are significantly different. We used the pairwise_test() function from the rstatix package (Kassambara, 2023). An ANOVA was conducted on three factors - sex, region, and the interaction between sex and region. After variance analyses, we conducted Tukey's Honestly Significant Difference (HSD), with the TukeyHSD function (Miller, 1981; Yandell, 1997), post-hoc tests on the paired means from our ANVOA results for the three factors above. To understand if there are random effects that different fly lines may have on the model, we used a Linear mixed model fit using Satterthwaite's method from the package lmerTest function lmer4() (Christensen & Kuznetsova) using fly lines as the random effect.

This study also recorded post-UV exposure activity. We conducted a MANOVA (Hand & Taylor, 1987; Krzanowski, 1988) to test differences between categories of fell status (active, barely active, inactive, dead) across activity time, fly region, and sex using the manova() function in R. To test the interaction of activity fell time (real-time UV exposure tolerance) to recovery (post-UV activity) we used a general linear model using the Satterthwaite's method for male and female flies, independently, to test the interaction of UV tolerance and recovery across region.

*Genomic insights*

To determine if there was a genomic basis underpinning the empirical UV exposure results, we used previously generated RNA sequence data for 5 lines of *Drosophila* (Turner et al. 2021). For each line we extracted the gonads and sequenced both the gonads and soma using 5 replicates for each sex and tissue. We obtained a list of known UV tolerance candidate genes from Svetec et al. 2016. In addition, we performed a genome-wide screen for expression changes of DNA repair, stress-tolerance, and UV tolerance genes.

To evaluate differential expression in the samples we used the program CuffDiff v2.2.2 (Trapnell et al. 2013). We identified statistically significant differential expression in the samples using Fisher's Adjusted P-values for each gene. These were generated using the P-values output by the initial analysis and correcting for any samples that had data at the same site. For any gene that showed significant differential expression in a sample, we conducted functional gene annotation analysis with DAVID v2023q1 (Huang et al. 2009). This yields a list of genes that had statistically significant differential expression, what function the gene has, and what the fold change was from CuffDiff.

## Results:

### *Solar Radiation and Distribution Map*

Estimates of higher solar radiation (kJ $m^{-2}$ $day^{-1}$) concur with the distribution of *D. santomea* (Fig. 1) at higher elevations and higher solar radiation. *D. santomea* also, potentially, inhabits certain areas along hill sides with low solar radiation. Lachaise et al. 2000 describes the habitat of *D. santomea* as a species that is found in mist forests, while *D. yakuba* is found in more disturbed areas (Llopart et al. 2005a). Current evidence suggests that *D. santomea* experiences higher UV exposure across its range. Unless currently unidentified physical solutions exist to shelter flies from UV, we expect radiation stress to be greater for this higher altitude species.

### *Apparatus function*

The function of developing this UV tolerance apparatus allows for reproducibility and dissemination for future experiments. Here we generated 1080p MP4 codec video files to observe the UV tolerance of flies from the island of São Tomé between island *D. yakuba* and *D. santomea* with mainland *D. yakuba* fly lines used as controls. Average temperature during trials ranged from 21.9°C-23.4°C (minimum temperature 19.7°C and a maximum temperature 29.7°C), reducing changes of heat shock during UV stress—a key improvement over prior work in UV crosslinkers which required heat shock surveys to separate effects from entangled variables (Matute & Harris, 2013).

The full set-up of the apparatus uses commonly found equipment in a biological laboratory. The apparatus set-up and enclosure construction can be found in our experiment in the methods. Our methods, assemblage, and construction can be modified to suit any lab space as needed. Given the enclosures are made from easily manipulatable acrylic, air-ports can be used to inject $CO_2$ directly into an enclosure if an experiment requires anesthetization before enclosures are pulled. Here the function of our innovative apparatus serves four purposes: containment, specimen recovery, digital recording, and a platform for interaction of an external force to test a phenotypic trait. We consider this apparatus to be novel in both its description and its use. The platform may be modified to a particular external phenomenon being addressed. This setup improves reproducibility and rigor in UV studies as videos are available for review, and UV exposure and temperature are finely controlled during exposure. Furthermore, remote filming allows for a much safer work environment, where the operator does not need to be in front of the apparatus potentially exposing themselves unnecessarily to ultraviolet radiation, unlike those found in the literature. To our knowledge, we have not seen an apparatus like the one used in this study (Fig. 2). Schematics are publicly available to facilitate similar studies in the field (Supplementary Material).

### *UV tolerance until inactivation*

Across all flies we observe variation in the UV tolerance among both the region flies reside as well as sex specific variation. There is a significant difference between sex for UV tolerance ([$P = 0.004$]; Fig. 3, Table 1). Males have a broader distribution of tolerance, while females have a tighter distribution and are more UV tolerant.

There are significant differences between UV tolerance of a species given the region it resides in ([P < 0.001]; Fig. 3, Table 1). At higher elevations the ultraviolet exposure is higher than at lower levels. A niche with higher elevation will have more UV stress than niche space at lower elevations (Fig. 1), and intrinsic UV tolerance matches well with habitat in nature. The mainland control lines of *D. yakuba* are not as UV tolerant as island resident *D. yakuba* or *D. santomea*. Tukey HSD test show that mainland *D. yakuba* is significantly different from both island *D. yakuba* ([P < 0.001]; for all comparisons) and *D. santomea* ([P < 0.001]; for all comparisons) but are not significantly different from each other (See Table 1 for all P-value comparisons). However, there are observable differences in both island *D. yakuba* and *D. santomea* in the distribution of when a fly becomes inactive through UV exposure (Fig. 3). Conditioning on sex, conducting a Wilcoxon test finds that there is a significant difference [P < 0.001] between each region for each sex independently, except for island *D. yakuba* and *D. santomea* males; [P = 0.451].

Among the interaction of sex and region there are significant differences between each factor and the interaction of the factors (Fig. 3, Table 1). Both sex and region are significantly different [P = 0.004 & P < 0.001] respectively, with the interaction between them being significantly different as well [P = 0.017]. HSD post-hoc test is used to determine which pair group means are significantly different from each other. We observe there is not a difference between sexes within the mainland *D. yakuba* [P = 0.345], *D. santomea* [P = 0978], or island *D. yakuba* [P = 0.071]. Between mainland *D. yakuba* and island *Drosophila* regions there is statistical difference between all interactions (Table 1). Conducting a linear mixed model fit using the Satterthwaite's method, there was no significant difference in post-UV or the interaction of post-UV time and region in female flies, but a significant difference in region. While male flies were significantly different in all factors and their interactions (Table 2). Males did not have fixed effects among the fly lines that are different from the model [intercept, P = 0.09403]. Females, conversely, have fixed effects given a fly line [intercept, P = 0.005] (Table 2).

*Recovery time after UV stress*

In natural systems, direct mortality from UV exposure is expected to be costly, but morbidity may affect survivorship and reproduction as well. The duration of inactivation would increase chances of predation in nature and reduce efforts for foraging. We find that being susceptible to UV radiation during active exposure assays may not mean recoverability is lower, especially given *D. santomea* is not melanistic like other *D. melanogaster* conspecifics (Wittkoop et al. 2003ab; Coolon et al. 2014; Svetec et al. 2016). UV recovery was observed across five-time intervals after UV tolerance filming. There is a higher density of island constituent species having higher number of active flies post-UV exposure compared to that of the mainland *D. yakuba* lines (Fig. 4). There are sex specific differences observed in female island *D. yakuba* ([P < 0.001]; Table 2) compared to the male island D. *yakuba* ([P = 0.191]; Table 2). Female island D. *yakuba* also have the highest survivorship among all regions ([P < 0.001]; Fig. 4, 5; Table 2). *D. santomea* has high survivorship among regions between sexes ([P < 0.001]; Fig. 4, 5; Table 2).

We observe a higher propensity for female flies to recover more rapidly and efficiently than males across regions. Among regions *D. santomea* has the highest recoverability. Across all groups by sex and region, female island *D. yakuba* have the highest recoverability ([P < 0.001]; Fig. 5, Table 3).

*Interactions Between Mortality versus Morbidity*

When observing if there is a difference between mortality vs morbidity, we conducted a MANOVA comparing states of recovery by post-UV time (time of fall during UV exposure), sex, and region. There is no significant difference between recovery state (response) and time of UV exposure. Among our other treatments there is a significant difference with sex and region and their interaction. There is no effect of when a fly faints from UV exposure to the state of recovery post-UV exposure, even with interactions with sex and region (Table 3). The *Pillai's* trace show that region a fly inhabits, as an independent variable, has the largest effect followed by the interaction of region and sex, then sex.

*Genomic Insights*

Understanding the genomic underpinnings of UV resistance and DNA repair is important to understand the adaptability and survivability at higher elevations where *D. santomea* resides. To identify genetic changes that might potentially influence UV tolerance, we performed differential expression testing. Expression analysis identifies several known UV tolerance genes or DNA repair genes. A list of significantly differentially expressed UV resistance genes, from DAVID analyses, and the strain the genes are listed in Table 4.

*Pp2B-14D* is upregulated in *D. santomea* males, while being down regulated in females, [♂ 1.30916 and ♀ -0.296757]. We find *CycG*, a meiotic recombination DNA repair gene (Nagel et al. 2012), is downregulated in island *D. yakuba*. The gene *Rtel1* (annotated as CG4078) is implicated in DNA repair, as well as the maintenance of genomic and male germ line stability (Yang et al. 2021). This gene is found to be significantly differentially expressed in *D. santomea*. In all lines, among sexes, it is downregulated, except in male line Thera6, where it is upregulated. *Syx13* is downregulated in all populations. DNA damage response gene *ctrip* (Gaudet et al. 2011) is downregulated in all *D. santomae* except Thera6. *Mei-9* is a known DNA nucleotide excision repair gene that is downregulated in only *D. santomea* Thera2005 [♀ -0.774595 and ♂ -1.60825]. *Mus304* has interactions with *mei-9* is only differentially expressed in *D. yakuba* mainland populations and is down regulated [♀ -0.830693 and ♂ -4.03084], especially in males.

The spellchecker1 (*spel1*) gene, involved in post-replication mismatch repair, is found across all populations but *D. santomea* line B1300.5. It is downregulated in all *D. yakuba* populations, except for female Cascade-1916. Where *spel1* is not significantly differentially expressed. However, it is found upregulated in female fly lines of *D. santomea* but is downregulated in males. We find that in *D. santomea* line Thera6, there is upregulation of transforming acidic coiled-coil protein (*tacc*) for female ovaries and male soma but downregulated in the testes. Finally, we found an uncategorized gene *CG5181* that is predicted to be involved in double stranded break repair (Barclay et al. 2014). This gene is predicted to be activated in response to ionizing radiation. It is exclusively found at significant upregulation in *D. santomea* line Thera6 [♀1.79065 and ♂ 1.5854]. In total, we identify 10 loci with gene expression changes across groups which are already known to influence UV tolerance.

## Discussion:

*Island Adaptation – Standing Variation and New Mutations*

Evolutionary innovation can come from standing variation or from new mutation (Hermisson & Pennings, 2005). Adaptation through standing variation is thought to occur quickly as phenotypic changes may be immediately beneficial under environmental changes (Goldschmidt, 1982). New mutation, on the other hand, requires longer time periods to facilitate evolutionary change (Smith, 1971; Hermisson & Pennings, 2005). In this study, we identify standing phenotypic variation for ultraviolet tolerance at low levels in mainland populations that would be immediately available to facilitate shifts to higher altitude. This pre-existing variation would facilitate immediate colonization of new ultraviolet radiation exposed niches, even while new mutations may have accumulated over time to improve resistance further. Such a scenario is consistent with Fisher's geometric model where key innovations of large initial effect are followed by multiple mutations that refine phenotypes further (Orr, 2005).

In our empirical testing we find there is a strong signal that UV tolerance is adapted on the island of São Tomé compared to the broadly distributed mainland Africa *D. yakuba*. This is not surprising given that island systems confer new endemics and novel traits throughout centuries of literature. Local adaptation to the island environment may come from new mutations or standing variation. The expectation is that standing variation allows for populations to adapt immediately, carrying traits over from the ancestral population that then fix in a new environment (Barrett & Schluter, 2008; Rogers et al. 2014). There is evidence that standing variation has a role in local adaptation. The distribution of UV tolerant mainland *D. yakuba* (ancestral population) is low in frequency; however, there is a segment of the population that is as UV tolerant as the island populations (Fig. 3). There are significant differences between populations of *D. santomea* and *D. yakuba* on the island compared to the mainland, but not between each other, i.e., shared on the island. This suggests that the UV tolerance is adaptable, and that at least some of this adaptability comes from standing variation.

However, our study may show too that new mutations have led to further adaptability than simply standing variation from mainland Africa. São Tomé *Drosophila* at higher altitudes shows higher tolerance to UV exposure than at lower altitudes. Given, too, that there is a strict geographic barrier where island *D. yakuba* do not inhabit but *D. santomea* thrive, and *vice versa* (Lachaise et al. 2000; Cariou et al. 2001; Llopart et al. 2005), illustrates that standing variation is likely not the only process of local adaptation (Assis & Bachtrog, 2013). Furthermore, the colonization of *D. yakuba* on the island is approximately 400,000 years later than *D. santomea*, thus giving time for new mutations to proliferate. Recoverability is much higher in *D. santomea* than in *D. yakuba*. Further implying that *D. santomea* is better adapted to tolerating UV radiation (Fig. 5). Previous genetic work has also found that *D. santomea* has higher *de novo* exons (Stewart & Rogers, 2019) and more fixed mutations not found in *D. yakuba* (Turner et al. 2021), leading us to suspect that novel genetic elements outside of standing variation has allowed for this observed local adaptation at higher elevations.

In island *D. yakuba*, populations that are at higher elevation do become darker than populations at lower elevations (Matute & Harris, 2013). All the while, higher elevation populations of *D. santomea* do not increase their pigmentation, continuing to have a lack of a melanistic trait. While temperature appears to be a competitive trait between the species (Comeault & Matute, 2021), from our work UV tolerance too appears to be competitive, and potentially even sex-specific. Especially given that female *D. yakuba* are highly UV tolerant and recover at the highest rate post-UV exposure (see section *Sex-specific Effects*). Given the curious case that *D. santomea* has reduced melanin, compared to other related *Drosophila*, there may be an intrinsic or extrinsic reason to how this species copes and combats the negative effects of ultraviolet radiation at higher elevations. To surmise, given that *D. santomea* has had longer isolation on the island and colonization to more specific niche space at higher elevations, it is not surprising that new mutations have a bigger role in local adaptation of the species. In comparison, standing variation appears to have a larger role in, the more recent, colonization of *D. yakuba* to the island where traits were fixed in the island population, brought over from the mainland.

*An Outlier to Altitude Induced Melanism?*

During the course of this experiment, we ran distribution models with MaxEnt 3.4.1 (Phillips et al. 2017) from locality data taken from Comeault & Matute, 2021. Without more robust sampling to point localities associated with samples, we, by no means, find our models anything more than speculative. Though, besides *D. santomea* being more UV tolerant than *D. yakuba*, annual geographic UV radiation data from WorldClim 2.1 potentially shows extrinsic sources for the lack of melanistic patterning at higher altitude. Lachaise et al. 2000 described the habitat of *D. santomea* residing in mist forests. These mist forests may have less UV exposure than disturbed areas. Our current evidence illustrates that *D. santomea* are in more UV exposed areas. However, more detailed sampling is required to elucidate if there is an extrinsic source to UV resistance and lack of melanistic qualities before surmising that *D. santomea* is an outlier to the current hypothesis of "altitudinal induced melanism."

*Sexual Dimorphism for UV tolerance*

Sex-specific phenotypes can free populations from evolutionary tradeoffs in reproductive strategies and behavioral risks between males and females (Darwin, 1871; Fisher, 1931; Lande, 1980; Parker, 1979). Decoupling phenotypic effects across sexes frees evolution from pleiotropic constraints to achieve sex-specific optima (Trivers, 1972; Lande, 1980; Albert & Otto, 2005). Our results are consistent with these models for evolution of dimorphic traits. We observe that sex has a large role on UV tolerance. Even our control population of mainland Africa *D. yakuba*, the females are more UV tolerant than to males; however, this result is not significant from statistical tests when conditioning with region, only significant when analyzing sex. We do not observe any differences of UV tolerance in real-time when comparing within region. There is a significant difference between the sexes within this population. Given that females do better as a whole and do better than males within the island *D. yakuba* population, there is likely a component that females possess making them more well adapted to UV exposure. This may be female size, being larger than males. Or potentially an intrinsic genetic trait leading to less susceptibility to the negative effects of UV.

When observing recovery rates across region and sex, overall *D. santomea* recovers at a higher rate with more active flies across time than the other populations. With the exception of female island *D. yakuba* which has the highest consistency of active flies across all groups tested (Fig. 5). There is sexual dimorphism between the sexes across *D. yakuba*, where males are darker than females (Llopart et al. 2002). This again leads to why and how females on the island are more UV tolerant and recover better than males in *D. yakuba*. Especially given that higher levels of melanin, i.e., pigmentation, will provide more protection to damaging UV radiation.

The genetic basis of change across groups is also consistent with these sex-specific effects as several loci are found on sex chromosomes or show sex-specific differences in gene expression changes. Pigmentation differences between *D. santomea* and *D. yakuba* are attributed to the X chromosome with nearly 90% of markers residing there (Llopart et al. 2002). Of our candidate DNA repair genes *Pp2B-14D*, *CG4078*, and *mei-9* all reside on the X chromosome and are only found in *D. santomea*. Both *Pp2B-14D* and *CG4078* are down regulated in the ovaries while also being upregulated in testes. Within female *D. santomea*, *spel1* is upregulated, but is downregulated within males and all *D. yakuba*, both island and mainland populations. Tissue-specific effects are expected occurrences, though being DNA repair genes and the dynamics observed where females have a lower mortality and morbidity than males, there may be pleiotropic effects, be them antagonistic or otherwise, that warrants further investigation. Especially given that the hybrid zone dynamics are asymmetric (Llopart et al. 2005) and our findings have females to be better adapted to ultraviolet radiation. These genetic factors may hold useful information to sex-specific effects on adaptation. We also find Sister-of-sex lethal (*ssx*) warrants further investigation in response to the *D. santomea – D. yakuba* island complex. Sister-of-sex lethal (*ssx*) is an inhibitor of sex lethal (*sxl*) auto-regulatory splicing (Moschall et al. 2019). *Ssx* is found in expression data from both island constituents, *D. santomea* and *D. yakuba*, but not in the mainland *D. yakuba*. Given these two species have a hybrid zone and are asymmetric in hybrid-cross viability, we recommend further investigation.

Previous sexual isolation studies conducted on both light and dark pigmentation conditions show no differences (Llopart et al. 2002), meaning that darkening is not driven by sexual selection. Instead, it is more likely an adaptable trait to higher UV exposure as population moves up in altitude (Pool & Aquadro, 2007). Lachaise et al. (2000) found that there is asymmetry in hybrids produced depending on cross. Here *D. santomea* follows Haldane's rule where the heterogametic sex is inviable (Coyne, 1985.) Our findings show that outside of potential regional adaptation, there are sex-specific effect to UV tolerance and recoverability. The reasons for these phenomena are only inferential in this study. However, our findings do illustrate that sex-specific effects do persist as it pertains to UV tolerance where females are more tolerant than males. The reason for this disparity requires further investigations.

*Genetic parallels across species*

UV tolerance has evolved in parallel in multiple populations and species of *Drosophila* (Pool & Aquadro, 2007; Zhao et al. 2015; Svetec et al. 2016). If few pathways exist to facilitate phenotypic change, we may expect to observe higher rates of convergence at the genetic level when we observe convergent changes across phenotypes (Stern, 2013). However, if there exist multiple independent genetic solutions, genetic convergence may be less common. As an initial step toward connecting genotype to phenotype, we identify multiple UV tolerance genes that show gene expression changes in *D. santomea*. Prior experiments have found many of the same genes or similar pathways identified in our own genetic analysis of UV resistance. The genes *phr*, *Dmp8*/*TTDA*, and *mei-9* as potential mutants that are involved in ultraviolet resistance, having strong photo-repair activity in the *Drosophila* Trichothiodystrophy Model (Boyd & Harris, 1987; Yildiz et al. 2004; Aguilar-Fuentes et al. 2008). Neither *phr* or *Dmp8*/*TTDA* are found to have any significant up/down regulation in *D. santomea* or *D. yakuba* on the island. However, *mei-9* is significantly downregulated in *D. santomea* in male and female gonads (Table 4). Potentially suggesting *Drosophila* of São Tomé uses a difference genomic mechanism for UV resistance and is a process of convergent evolution for local adaptation to ultraviolet radiation compared to *D. melanogaster*.

There is potential for parallel changes in gene expression in comparison with other systems. Our functional analysis has overlapping genes with the UV gene list from Svetec et al. (2016) at *CycG*, *mei-9*, *mus304*, and *spel1*. In this study Syntaxin 13 (*Syx13*) is significantly downregulated across all populations, with all using parallel gene expression. Turner et al. (2021) found three more candidate genes that have significant rearrangements that are annotated as UV resistance genes which include *Parp*, *Victoria*, and *spel1*. Both *Parp* and *spel1* have significant expression changes in *D. santomea*. These findings allow for hints as to how *D. santomea* can cope with ultraviolet radiation without a mechanistic response, e.g., producing more melanin, unlike other organisms with ranges at high altitudes. Grapes (*grp*) is another DNA repair gene that is found only in the *D. santomea* group from our DAVID analysis.

Outside of shielding oneself from UV rays, one must hypothesize that an intrinsic trait must be able to combat this extrinsic phenomenon. Given our assays on expression data, we find that this hypothesis is worth further investigation with future genomic analysis.

*Conclusion*

Throughout this study we have elucidated that, of the populations surveyed, *D. santomea* residing at higher elevations, thus exposed to more ultraviolet radiation, do indeed tolerate UV better than *D. yakuba*. Island *D. yakuba* populations tolerate ultraviolet radiation better than those of their mainland constituents. Sex-specific effects, too, play a role in UV tolerance, where females have a higher tolerance to UV radiation than males in all populations. Female island *D. yakuba* are even more UV tolerant than *D. santomea*, both on male and female sexes curiously. Our study also produces an easily reproducible UV exposure apparatus with commonly found lab equipment, with provided enclosure schematics for other researchers to use for future experiments (Supplementary Material). Future work should focus on the genomic underpinnings of UV tolerance within the system studied here. It is an abnormal system that does not follow current observations of altitude-induced melanism. Meaning that further research should be conducted on the genes potentially allowing for less pigmented *D. santomea* to occur at higher altitudes than other *Drosophila*. Especially given that this research shows that, holistically, *D. santomea* are empirically more UV tolerant than their more melanistic conspecifics.

## Acknowledgements:
We would like to acknowledge Dr. Daniel Matute for graciously supplying *Drosophila* lines of both species from the island of São Tomé. We too need to acknowledge the National *Drosophila* Species Stock Center at Cornell College of Agriculture and Life Sciences for providing mainland control lines. We would also like to thank Dr. Farnaz Fouladi and Dr. Gabe O'Reilly for their knowledge and insights into statistical analyses and accuracy.

**Funding:**
This work was funded by NIH NIGMS MIRA R35-GM133376 under Rebekah L. Rogers.

Conflicted of interest: There are no conflicts of interest to declare.

Ethical Approval: Not applicable.

**References:**

Aguilar-Fuentes, J., M.Fregoso, M.Herrera, E.Reynaud, C.Braun, J. E.Egly, and M.Zurita. 2008. p8/TTDA overexpression enhances UV-irradiation resistance and suppresses TFIIH mutations in a Drosophila trichothiodystrophy model. PLoS Genet. 4:e1000253.

Albert, A. Y., & Otto, S. P. (2005). Sexual selection can resolve sex-linked sexual antagonism. Science, 310(5745), 119-121.

Assis, R., & Bachtrog, D. (2013). Neofunctionalization of young duplicate genes in Drosophila. Proceedings of the National Academy of Sciences, 110(43), 17409-17414.

Barclay, S. S., Tamura, T., Ito, H., Fujita, K., Tagawa, K., Shimamura, T., ... & Okazawa, H. (2014). Systems biology analysis of Drosophila in vivo screen data elucidates core networks for DNA damage repair in SCA1. *Human molecular genetics*, *23*(5), 1345-1364.

Barrett, R. D., & Schluter, D. (2008). Adaptation from standing genetic variation. Trends in ecology & evolution, 23(1), 38-44.

Blyth, J. E., Lachaise, D., & Ritchie, M. G. (2008). Divergence in multiple courtship song traits between Drosophila santomea and D. yakuba. *Ethology*, *114*(7), 728-736.

Bonduriansky, R., & Chenoweth, S. F. (2009). Intralocus sexual conflict. Trends in ecology & evolution, 24(5), 280-288.

Brown, R. M., Siler, C. D., Oliveros, C. H., Esselstyn, J. A., Diesmos, A. C., Hosner, P. A., ... & Alcala, A. C. (2013). Evolutionary processes of diversification in a model island archipelago. Annual Review of Ecology, Evolution, and Systematics, 44, 411-435.

Cande, J., Andolfatto, P., Prud'homme, B., Stern, D. L., & Gompel, N. (2012). Evolution of multiple additive loci caused divergence between Drosophila yakuba and D. santomea in wing rowing during male courtship.

Bauer, D. F. (1972). Constructing confidence sets using rank statistics. *Journal of the American Statistical Association* **67**, 687–690. doi:10.1080/01621459.1972.10481279.

Darwin, C. (1859). On the origins of species by means of natural selection. London: Murray, 247, 1859.

Darwin, C. (1871). *The descent of man, and selection in relation to sex*.

de Souza, A. R., Mayorquin, A. Z., & Sarmiento, C. E. (2020). Paper wasps are darker at high elevation. Journal of Thermal Biology, 89, 102535.

Dobzhansky, T. (1937). Genetics and the Origin of Species. Columbia University Biological Series. New York: Columbia University Press. LCCN 37033383. OCLC 766405.

Boyd, J. B., & Harris, P. V. (1987). Isolation and characterization of a photorepair-deficient mutant in Drosophila melanogaster. *Genetics*, *116*(2), 233-239.

Brenner, M., & Hearing, V. J. (2008). The protective role of melanin against UV damage in human skin. *Photochemistry and photobiology*, *84*(3), 539-549.

Cariou, M. L., Silvain, J. F., Daubin, V., Da Lage, J. L., & Lachaise, D. (2001). Divergence between Drosophila santomea and allopatric or sympatric populations of D. yakuba using paralogous amylase genes and migration scenarios along the Cameroon volcanic line. *Molecular Ecology*, *10*(3), 649-660.

Christensen, R. H. B. & Kuznetsova, A. for the overload in **lmerTest** – **lme4**-authors for the underlying implementation in **lme4**.

Comeault, A. A., & Matute, D. R. (2021). Temperature-dependent competitive outcomes between the fruit flies Drosophila santomea and Drosophila yakuba. The American Naturalist, 197(3), 312-323.

Coolon, J. D., McManus, C. J., Stevenson, K. R., Graveley, B. R., & Wittkopp, P. J. (2014). Tempo and mode of regulatory evolution in Drosophila. Genome research, 24(5), 797-808.

Coyne, J. A. (1985). The genetic basis of Haldane's rule. Nature, 314(6013), 736-738.

Coyne, J. A., Kim, S. Y., Chang, A. S., Lachaise, D., & Elwyn, S. (2002). Sexual isolation between two sibling species with overlapping ranges: Drosophila santomea and Drosophila yakuba. *Evolution*, *56*(12), 2424-2434.

Coyne, J. A., & Orr, H. A. (2004). Speciation Sinauer Associates. Sunderland, MA, 276, 281.

Coyne, J. A., Elwyn, S., & Rolan-Alvarez, E. (2005). Sexual isolation between Drosophila yakuba and D. santomea: effects of environment and experimental design. Evolution, 59, 2588-2601.

de Souza, A. R., Mayorquin, A. Z., & Sarmiento, C. E. (2020). Paper wasps are darker at high elevation. *Journal of Thermal Biology*, *89*, 102535.

Emery, N. J., & Clayton, N. S. (2004). The mentality of crows: convergent evolution of intelligence in corvids and apes. science, 306(5703), 1903-1907.

Fick, S. E., & Hijmans, R. J. (2017). WorldClim 2: new 1-km spatial resolution climate surfaces for global land areas. International journal of climatology, 37(12), 4302-4315.

Fisher, R. A. (1931). 093: The Evolution of Dominance.

Gaudet, P., Livstone, M. S., Lewis, S. E., & Thomas, P. D. (2011). Phylogenetic-based propagation of functional annotations within the Gene Ontology consortium. *Briefings in bioinformatics*, *12*(5), 449-462.

Goldman, E. A. (1947). The pocket gophers (genus Thomomys) of Arizona. *North American Fauna*, (59 (59)), 1-39.

Goldschmidt, R. (1982). *The material basis of evolution* (Vol. 28). Yale University Press.

Haldane, J. B. (1922). Sex ratio and unisexual sterility in hybrid animals. Journal of genetics, 12, 101-109.

Hand, D. J. and Taylor, C. C. (1987) Multivariate Analysis of Variance and Repeated Measures. Chapman and Hall.

Harris, R. M., McQuillan, P., & Hughes, L. (2013). A test of the thermal melanism hypothesis in the wingless grass-hopper Phaulacridium vittatum. *Journal of Insect Science*, *13*(1), 51.

Hermisson, J., & Pennings, P. S. (2005). Soft sweeps: molecular population genetics of adaptation from standing genetic variation. Genetics, 169(4), 2335-2352.

Hollander, M. & Douglas A. Wolfe (1973). *Nonparametric Statistical Methods*. New York: John Wiley & Sons. Pages 27–33 (one-sample), 68–75 (two-sample). Or second edition (1999).

Huang, D. W., Sherman, B. T., Lempicki, R. A., Huang, D. W., & Sherman, B. T. L. R. A. (2009). DAVID Functional Annotation Bioinformatics Microarray Analysis. Nat Protoc, 4, 44-57.

Kassambara A (2023). _rstatix: Pipe-Friendly Framework for Basic Statistical Tests_. R package version 0.7.2, <https://CRAN.R-project.org/package=rstatix>.

Kauffman, S., & Levin, S. (1987). Towards a general theory of adaptive walks on rugged landscapes. Journal of theoretical Biology, 128(1), 11-45.

Krzanowski, W. J. (1988) *Principles of Multivariate Analysis. A User's Perspective.* Oxford.

Lachaise, D., Harry, M., Solignac, M., Lemeunier, F., Benassi, V., & Cariou, M. L. (2000). Evolutionary novelties in islands: Drosophila santomea, a new melanogaster sister species from Sao Tome. *Proceedings of the Royal Society of London. Series B: Biological Sciences*, *267*(1452), 1487-1495.

Lande, R. (1980). Sexual dimorphism, sexual selection, and adaptation in polygenic characters. Evolution, 292-305.

Lee, S. K. (2018). Sex as an important biological variable in biomedical research. BMB reports, 51(4), 167.

Llopart, A., Elwyn, S., Lachaise, D., & Coyne, J. A. (2002). Genetics of a difference in pigmentation between Drosophila yakuba and Drosophila santomea. *Evolution*, *56*(11), 2262-2277.

Llopart, A., Lachaise, D., & Coyne, J. A. (2005a). An anomalous hybrid zone in Drosophila. Evolution, 59(12), 2602-2607.

Llopart, A., Lachaise, D., & Coyne, J. A. (2005b). Multilocus analysis of introgression between two sympatric sister species of Drosophila: Drosophila yakuba and D. santomea. Genetics, 171(1), 197-210.

Llopart, A. (2012). The rapid evolution of X-linked male-biased gene expression and the large-X effect in Drosophila yakuba, D. santomea, and their hybrids. Molecular biology and evolution, 29(12), 3873-3886.

Matute, D. R., Novak, C. J., & Coyne, J. A. (2009). Temperature-based extrinsic reproductive isolation in two species of Drosophila. *Evolution*, *63*(3), 595-612.

Matute, D. R., & Harris, A. (2013). The influence of abdominal pigmentation on desiccation and ultraviolet resistance in two species of Drosophila. *Evolution*, *67*(8), 2451-2460.

Miller, R. G. (1981) Simultaneous Statistical Inference. Springer.

Moehring, A. J., Llopart, A., Elwyn, S., Coyne, J. A., & Mackay, T. F. (2006a). The genetic basis of prezygotic reproductive isolation between Drosophila santomea and D. yakuba due to mating preference. *Genetics*, *173*(1), 215-223.

Moehring, A. J., Llopart, A., Elwyn, S., Coyne, J. A., & Mackay, T. F. (2006.). The genetic basis of postzygotic reproductive isolation between Drosophila santomea and D. yakuba due to hybrid male sterility. *Genetics*, *173*(1), 225-233.

Moschall, R., Rass, M., Rossbach, O., Lehmann, G., Kullmann, L., Eichner, N., ... & Medenbach, J. (2019). Drosophila Sister-of-Sex-lethal reinforces a male-specific gene expression pattern by controlling Sex-lethal alternative splicing. Nucleic acids research, 47(5), 2276-2288.

Nagel, A. C., Fischer, P., Szawinski, J., La Rosa, M. K., & Preiss, A. (2012). Cyclin G is involved in meiotic recombination repair in Drosophila melanogaster. *Journal of Cell Science*, *125*(22), 5555-5563.

Obbard, D. J., Maclennan, J., Kim, K. W., Rambaut, A., O'Grady, P. M., & Jiggins, F. M. (2012). Estimating divergence dates and substitution rates in the Drosophila phylogeny. *Molecular biology and evolution*, *29*(11), 3459-3473.

Orr, H. A. (2005). The genetic theory of adaptation: a brief history. Nature Reviews Genetics, 6(2), 119-127.

Parker, G. A. (1979). Sexual selection and sexual conflict. Sexual selection and reproductive competition in insects, 123, 166.

Peluffo, A. E., Hamdani, M., Vargas-Valderrama, A., David, J. R., Mallard, F., Graner, F., & Courtier-Orgogozo, V. (2021). A morphological trait involved in reproductive isolation between Drosophila sister species is sensitive to temperature. Ecology and Evolution, 11(12), 7492-7506.

Phillips, S. J., Anderson, R. P., Dudík, M., Schapire, R. E., & Blair, M. E. (2017). Opening the black box: An open-source release of Maxent. Ecography, 40(7), 887-893.

Pool, J. E., & Aquadro, C. F. (2007). The genetic basis of adaptive pigmentation variation in Drosophila melanogaster. *Molecular ecology*, *16*(14), 2844-2851.

R Core Team (2022). R: A language and environment for statistical computing. R Foundation for Statistical Computing, Vienna, Austria. URL https://www.R-project.org/.

Reguera, S., Zamora-Camacho, F. J., & Moreno-Rueda, G. (2014). The lizard Psammodromus algirus (Squamata: Lacertidae) is darker at high altitudes. *Biological Journal of the Linnean Society*, *112*(1), 132-141.

Rice, W. R. (1987). The accumulation of sexually antagonistic genes as a selective agent promoting the evolution of reduced recombination between primitive sex chromosomes. Evolution, 41(4), 911-914.

Rogers, R. L., Cridland, J. M., Shao, L., Hu, T. T., Andolfatto, P., & Thornton, K. R. (2014). Landscape of standing variation for tandem duplications in Drosophila yakuba and Drosophila simulans. Molecular biology and evolution, 31(7), 1750-1766.

Rosenblum, E. B., Parent, C. E., & Brandt, E. E. (2014). The molecular basis of phenotypic convergence. Annual Review of Ecology, Evolution, and Systematics, 45, 203-226.

Rowe, L., Chenoweth, S. F., & Agrawal, A. F. (2018). The genomics of sexual conflict. The American Naturalist, 192(2), 274-286.

RStudio Team (2020). RStudio: Integrated Development for R. RStudio, PBC, Boston, MA URL http://www.rstudio.com/.

Schoville, S. D., Bonin, A., François, O., Lobreaux, S., Melodelima, C., & Manel, S. (2012). Adaptive genetic variation on the landscape: methods and cases. Annual Review of Ecology, Evolution, and Systematics, 43, 23-43.

Smith, J. M. (1971). What use is sex?. Journal of theoretical biology, 30(2), 319-335.

Stern, D. L. (2013). The genetic causes of convergent evolution. Nature Reviews Genetics, 14(11), 751-764.

Svetec, N., Cridland, J. M., Zhao, L., & Begun, D. J. (2016). The adaptive significance of natural genetic variation in the DNA damage response of Drosophila melanogaster. *PLoS genetics*, *12*(3), e1005869.

Trapnell, C., Hendrickson, D. G., Sauvageau, M., Goff, L., Rinn, J. L., & Pachter, L. (2013). Differential analysis of gene regulation at transcript resolution with RNA-seq. Nature biotechnology, 31(1), 46-53.

Trivers, R. (1972). Parental investment and sexual selection (Vol. 136, p. 179). Cambridge, MA: Biological Laboratories, Harvard University.

Turner, B. A., Miorin, T. R., Stewart, N. B., Reid, R. W., Moore, C. C., & Rogers, R. L. (2021). Chromosomal rearrangements as a source of local adaptation in island Drosophila. *arXiv preprint arXiv:2109.09801*.

Ulbing, C. K., Muuse, J. M., & Miner, B. E. (2019). Melanism protects alpine zooplankton from DNA damage caused by ultraviolet radiation. *Proceedings of the Royal Society B*, *286*(1914), 20192075.

Watson, E. T., Rodewald, E., & Coyne, J. A. (2007). The courtship song of Drosophila santomea and a comparison to its sister species D. yakuba (Diptera: Drosophilidae). *European Journal of Entomology*, *104*(1), 145.

Wittkopp, P. J., Williams, B. L., Selegue, J. E., & Carroll, S. B. (2003a). Drosophila pigmentation evolution: divergent genotypes underlying convergent phenotypes. Proceedings of the National Academy of Sciences, 100(4), 1808-1813.

Wittkopp, P. J., Carroll, S. B., & Kopp, A. (2003b). Evolution in black and white: genetic control of pigment patterns in Drosophila. TRENDS in Genetics, 19(9), 495-504.

Whittall, J. B., Voelckel, C., Kliebenstein, D. J., & Hodges, S. A. (2006). Convergence, constraint and the role of gene expression during adaptive radiation: floral anthocyanins in Aquilegia. Molecular Ecology, 15(14), 4645-4657.

Yandell, B. S. (1997) Practical Data Analysis for Designed Experiments. Chapman & Hall.

Yang, Y., Kong, R., Goh, F. G., Somers, W. G., Hime, G. R., Li, Z., & Cai, Y. (2021). dRTEL1 is essential for the maintenance of Drosophila male germline stem cells. *PLoS Genetics*, *17*(10), e1009834.

Yildiz O., H.Kearney, B. C.Kramer, and J. J.Sekelsky. 2004. Mutational analysis of the Drosophila DNA repair and recombination gene mei-9. Genetics 167:263–273.

Zhao, L., Wit, J., Svetec, N., & Begun, D. J. (2015). Parallel gene expression differences between low and high latitude populations of Drosophila melanogaster and D. simulans. *PLoS genetics*, *11*(5), e1005184.

Zhou, Z., Clarke, J., & Zhang, F. (2008). Insight into diversity, body size and morphological evolution from the largest Early Cretaceous enantiornithine bird. Journal of Anatomy, 212(5), 565-577.

Figures:

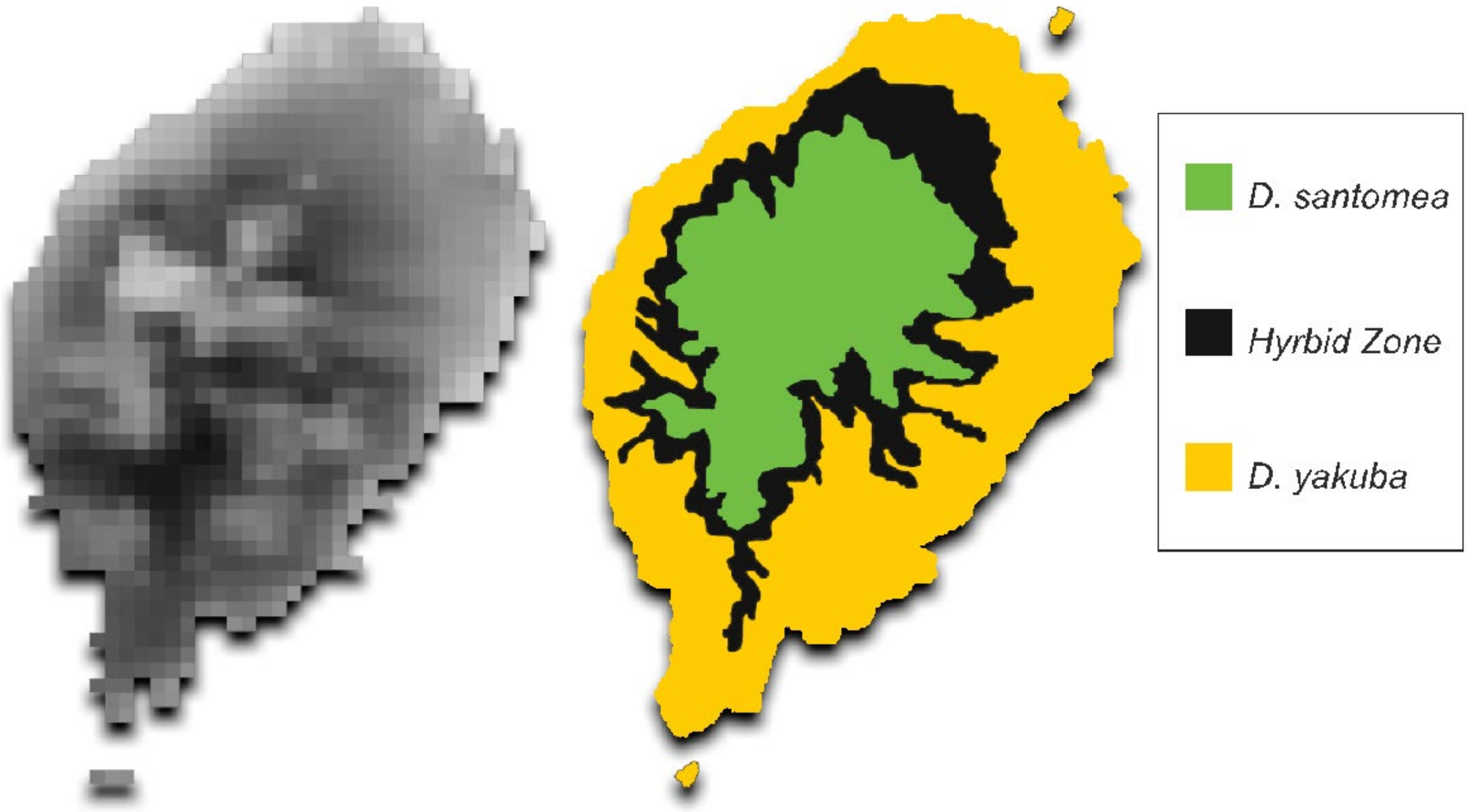

Fig. 1: Maps of the island of São Tomé. Left map is solar radiation data for WorldClim 2.1 historical climate data. The left map is an overlay among all months for a holistic annual solar radiation São Tomé is exposed to with lighter colors signifying higher solar radiation (kJ $m^{-2}$ $day^{-1}$) and darker colors delineating less solar radiation. The right map gives the estimated ranges between *D. santomea* and *D. yakuba* adapted from Lachaise et al. 2000).

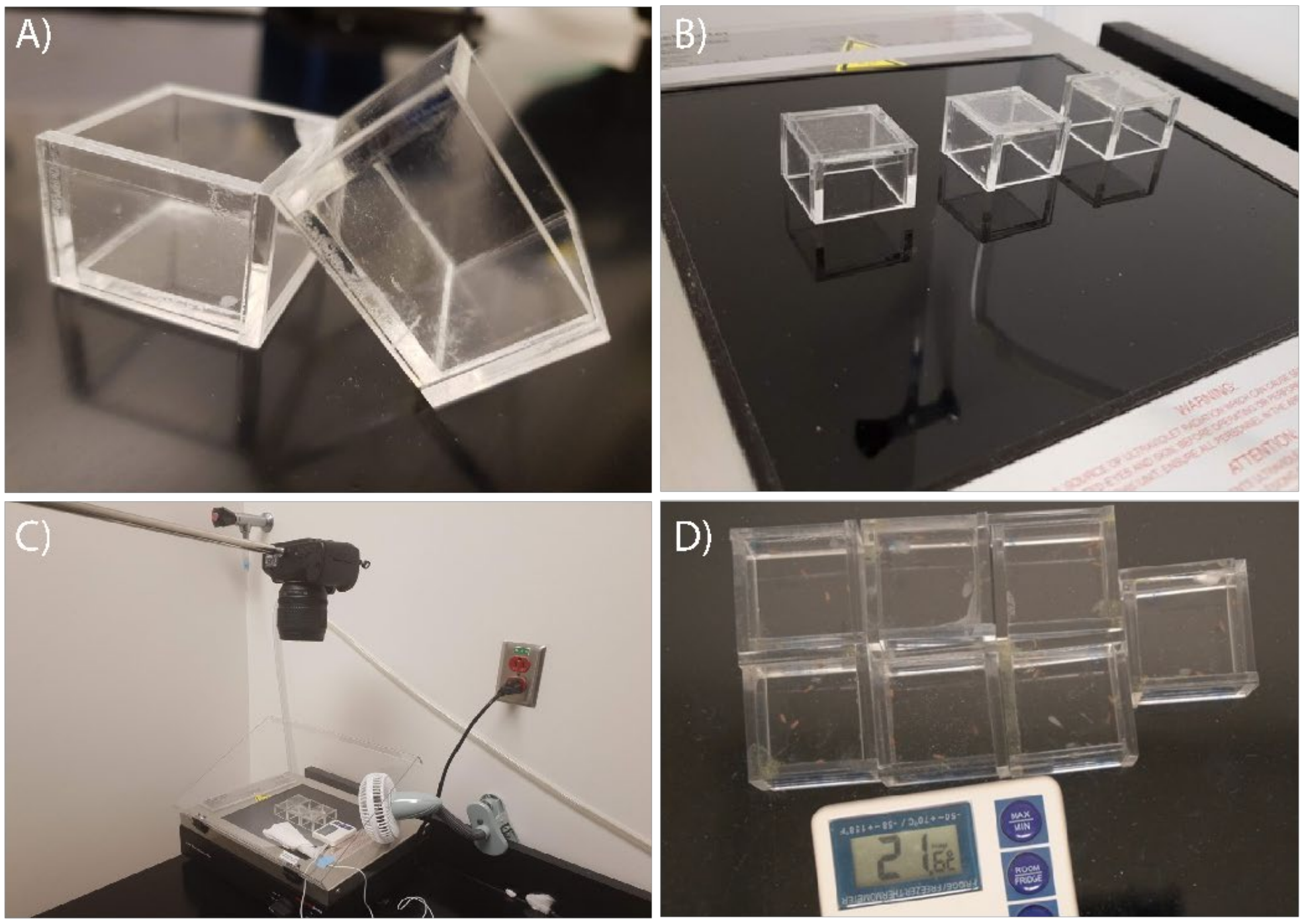

Figure 2: Recording apparatus set-up. A) Fully assembled enclosures to contain fly strain replicates. B) Enclosure placement on the UV platform. C) Full set-up of apparatus with cooling, temperature probe, camera, and enclosure placement. D) Field of view (FOV) with flies contained on UV transilluminator platform with temperature probe read out visible in an active experiment.

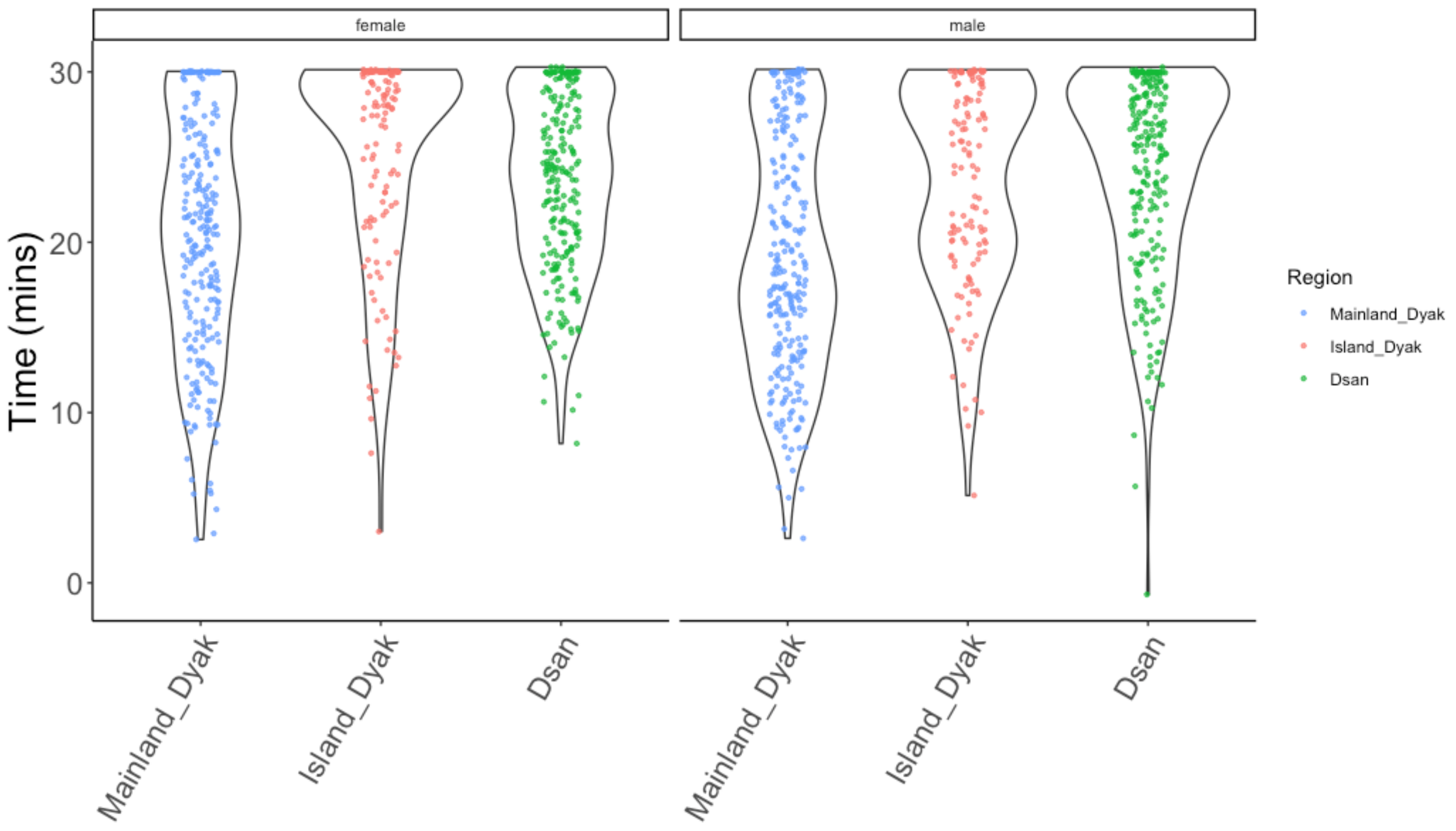

Figure 3: Box plot of all flies separated by sex-region. Females are the on the left with males on the right. Region is separated by label on the x-axis and by color with Mainland *D.yakuba*, *D. yakuba*, and *D. santomea* in light blue, orange-red, and green, respectively.

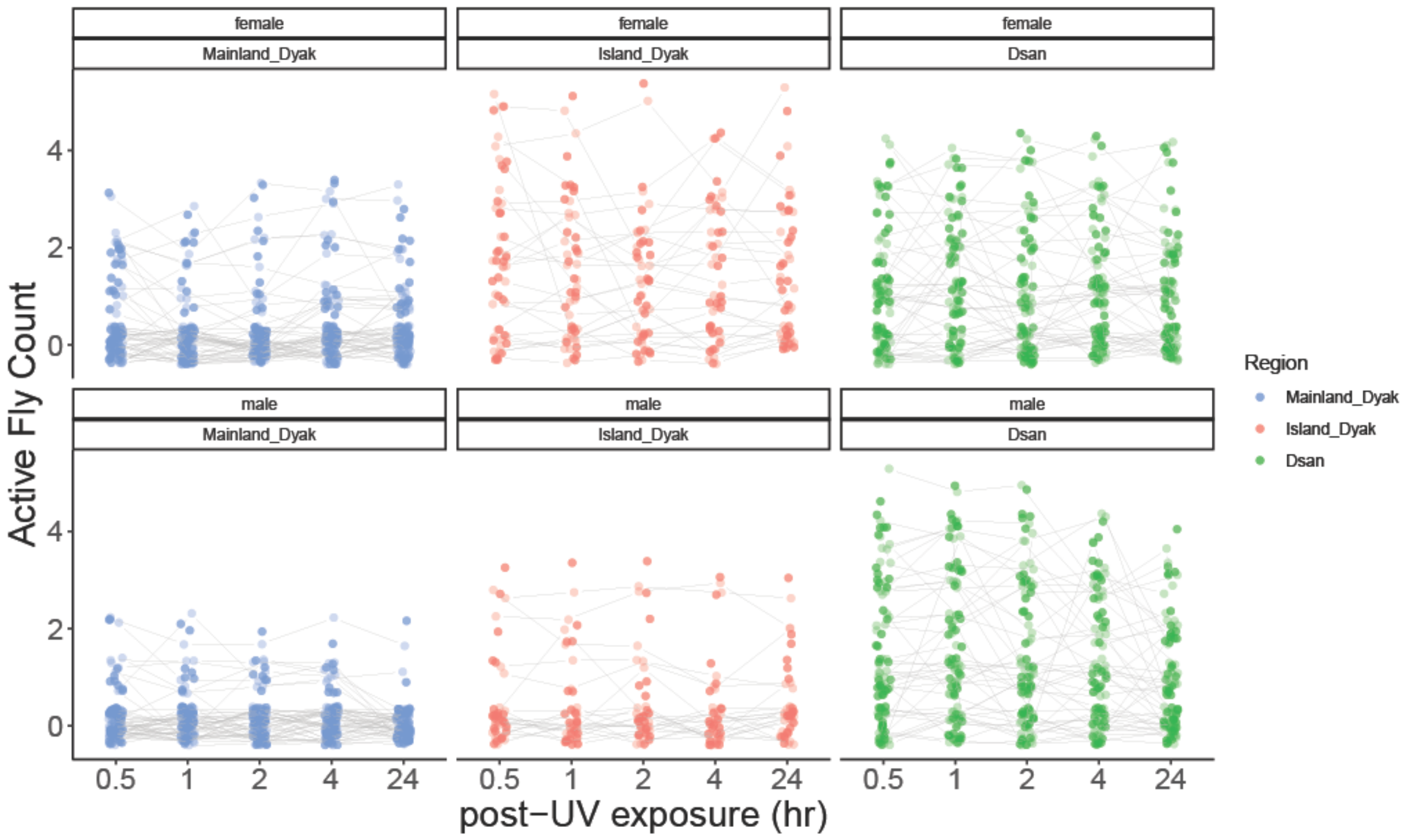

Figure 4: Split line plots of flies that are categorized as actively recovered partitioned by region and sex. Region is separated by label on the x-axis and by color with Mainland *D. yakuba*, *D. yakuba*, and *D. santomea* in light blue, orange-red, and green, respectively. Each dot signifies a fly vial with the number of flies in the *Active* category on the y-axis across each observed time point after UV screening.

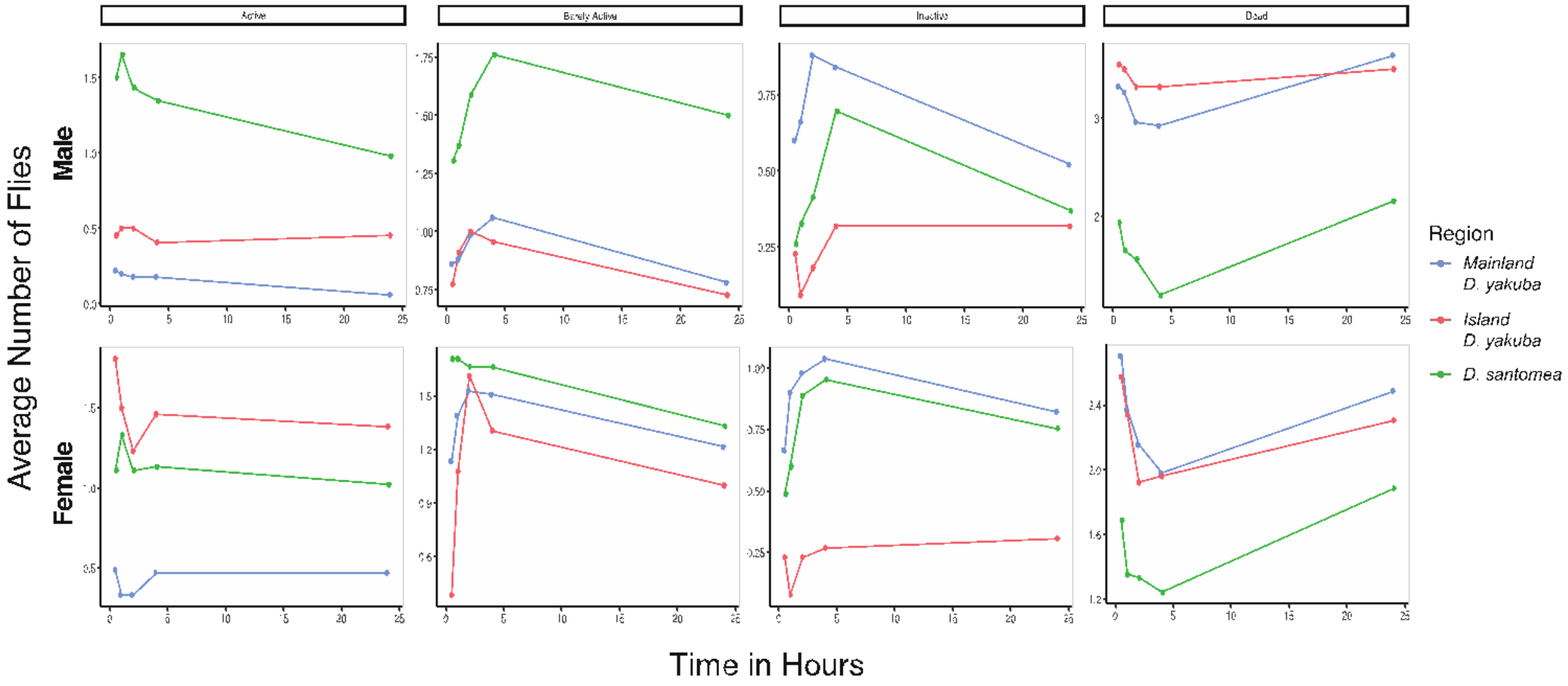

Figure 5: Line plots between A) males and B) females split by color of region with Mainland *D. yakuba*, *D. yakuba*, and *D. santomea* in light blue, orange-red, and green, respectively. The average number of flies on for each time intervals post-UV exposure on the y-axis. The x-axis is time in hours from 30 minutes to 24 hours post-UV exposure.

**Tables:**

Table 1: ANOVA of *Drosophila* sex and regions with the Tukey Honestly Significant Difference (HSD) test between sex-region specific differences statistics on real-time UV exposure. Significant code: P = 0 delineated ***.

| **ANOVA:** | by Sex-Region | | | | |
|---|---|---|---|---|---|
| | DF | Sum of Squares | Mean Sum of Squares | *F* | *P* |
| Sex | 1 | 132 | 131.8 | 8.199 | 0.00426 |
| Region | 2 | 3812 | 1906.1 | 118.588 | *** |
| Sex:Region | 2 | 131 | 65.6 | 4.079 | 0.01716 |
| Residuals | 1194 | 19191 | 16.1 | | |

| **TukeyHSD:** | by Sex-Region | | | |
|---|---|---|---|---|
| | Difference | Lower | Upper | *P-adj* |
| Male:Mainland Dyak-Female:Mainland Dyak | 0.704866 | -1.7127061 | 0.30297411 | 0.3449818 |
| Female:Island Dyak-Female:Mainland Dyak | 4.1763135 | 2.9559909 | 5.39663605 | *** |
| Male:Island Dyak-Female:Mainland Dyak | 2.822634 | 1.5862518 | 4.0590162 | *** |
| Female:Dsan-Female:Mainland Dyak | 3.2173747 | 2.181647 | 4.25310177 | *** |
| Male:Dsan-Female:Mainland Dyak | 3.4907644 | 2.4610355 | 4.52049334 | *** |
| Female:Island Dyak-Male:Mainland Dyak | 4.8811795 | 3.6567433 | 6.1056157 | *** |
| Male:Island Dyak-Male:Mainland Dyak | 3.5275 | 2.2870574 | 4.76794259 | *** |
| Female:Dsan-Male:Mainland Dyak | 3.9222407 | 2.8816701 | 4.96281143 | *** |
| Male:Dsan-Male:Mainland Dyak | 4.1956304 | 3.1610298 | 5.23023108 | *** |
| Male:Island Dyak-Female:Island Dyak | 1.3536795 | -2.7721799 | 0.06482089 | 0.0713222 |
| Female:Dsan-Female:Island Dyak | -0.9589387 | -2.2064294 | 0.28855188 | 0.2411661 |
| Male:Dsan-Female:Island Dyak | -0.6855491 | -1.9280642 | 0.55696613 | 0.6153994 |
| Female:Dsan-Male:Island Dyak | 0.3947407 | -0.8684642 | 1.65794565 | 0.9484894 |
| Male:Dsan-Male:Island Dyak | 0.6681304 | -0.5901612 | 1.92642204 | 0.6542787 |
| Male:Dsan-Female:Dsan | 0.2733897 | -0.7883953 | 1.33517465 | 0.9775918 |

Table 2: Type III ANOVA table conditioned by sex with random effects on fly lines with active flies post-UV exposure. Significant code: P = 0 delineated ***.

*Type III Analysis of Variance Table with Satterthwaite's method*

**Males**

Random effects:

| Groups | Name | Variance | Std. Dev. |
|---|---|---|---|
| Fly Lines | (Intercept) | 0.6633 | 0.8144 |
| Residual | | 0.2956 | 0.5437 |

Number of obs: 605, groups: experiment, 119

Fixed Effects:

| | Estimate | Std. Error | df | t-value | Pr(>\|t\|) |
|---|---|---|---|---|---|
| (Intercept) | 0.206893 | 0.122626 | 126.335644 | 1.687 | 0.09403 |
| Post-UV Time | -0.006174 | 0.00385 | 483.531688 | -1.603 | 0.10951 |
| Island *D. yak* | 0.27911 | 0.212395 | 126.335645 | 1.314 | 0.19119 |
| *D. san* | 1.299905 | 0.179064 | 125.852456 | 7.259 | *** |
| Post-UV Time:Region - Island Dyak | 0.005221 | 0.006669 | 483.531687 | 0.783 | 0.43411 |
| Post-UV Time:Region - Dsan | -0.015696 | 0.005562 | 483.531687 | -2.822 | ** |

*Type III Analysis of Variance Table with Satterthwaite's method*

**Females**

Random effects:

| Groups | Name | Variance | Std. Dev. |
|---|---|---|---|
| Fly Lines | (Intercept) | 0.9401 | 0.9696 |

| | | |
|---|---|---|
| Residual | 0.3535 | 0.5946 |

Number of obs: 610, groups: experiment, 121

| Fixed Effects: | | | | | |
|---|---|---|---|---|---|
| | Estimate | Std. Error | df | t-value | Pr(>\|t\|) |
| (Intercept) | 0.400589 | 0.143213 | 126.32144 | 2.797 | 0.005963 |
| Post-UV Time | 0.003019 | 0.004169 | 485.929963 | 0.724 | 0.469347 |
| Island *D. yak* | 1.117103 | 0.246457 | 126.32144 | 4.533 | *** |
| *D. san* | 0.782448 | 0.210347 | 126.110165 | 3.72 | *** |
| Post-UV Time:Region - Island Dyak | -0.00949 | 0.007175 | 485.929963 | -1.323 | 0.186541 |
| Post-UV Time:Region - Dsan | -0.010045 | 0.006089 | 485.929963 | -1.65 | 0.099652 |

Table 3: MANOVA results of recovery compared to sex, region, post-UV time, and their interactions. Significant code: P = 0 delineated ***.

| **MANOVA:** | | | | | | |
|---|---|---|---|---|---|---|
| | DF | Pillai | Approx. *F* | # DF | Density DF | *P* |
| Post-UV Time | 1 | 0.007508 | 2.27 | 4 | 1200 | 0.05983 |
| Region | 2 | 0.256329 | 44.138 | 8 | 2402 | *** |
| Sex | 1 | 0.04711 | 14.832 | 4 | 1200 | *** |
| Post-UV Time:Region | 2 | 0.004941 | 0.744 | 8 | 2402 | 0.65289 |
| Post-UV Time:Sex | 1 | 0.001987 | 0.97 | 4 | 1200 | 0.66457 |
| Region:Sex | 2 | 0.054333 | 8.385 | 8 | 2402 | *** |
| Post-UV Time:Region:Sex | 2 | 0.002146 | 0.323 | 8 | 2402 | 0.95779 |
| Residuals | 1203 | | | | | |

Table 4: Table of candidate genes from our RNA-seq genomic assays that have significant differential expression. This is list of genes is from a DAVID functional clustering analysis focusing on DNA repair mechanisms.

| GE Number | Gene Name | Abbr. Description | Line - Tissue | Log2 Fold Change |
|---|---|---|---|---|
| GE23311 | *CycG* | Cyclin G (CycG) | Island Cascade-1916 - Testes | -0.483561 |
| GE15960 | *Pp2B-14D* | Protein phosphatase 2B at 14D (Pp2B-14D) | Thera2005 - Ovaries | -0.296757 |
| | | | Thera2005 - Testes | 1.30916 |
| GE16425 | *CG4078* | Regulator of telomere elongation helicase 1 (Rtel1) | Thera6 - Ovaries | -0.341154 |
| | | | Thera6 - Testes | 0.403238 |
| | | | Thera2005 - Ovaries | -0.0300314 |
| | | | Thera2005 - Testes | -1.69884 |
| | | | B13005 - Ovaries | -0.312243 |
| | | | OBAT-12003 - Ovaries | -0.439944 |
| | | | OBAT-12003 - Testes | -1.3714 |
| GE21949 | *Syx13* | Syntaxin 13 (Syx13) | Thera6 - Ovaries | -0.700132 |
| | | | Thera6 - Testes | -0.793291 |
| | | | Island Cascade-1916 - Ovaries | -0.105177 |
| | | | Mainland Tai6 - Ovaries | -0.57351 |
| | | | Mainland Tai6 - Female soma | -2.29854 |
| | | | B13005 - Ovaries | -0.391115 |
| GE25331 | *ctrip* | circadian trip(ctrip) | Thera6 - Ovaries | 0.0313342 |
| | | | Thera6 - Testes | 0.0978823 |
| | | | Thera2005 - Ovaries | -0.823872 |
| | | | Thera2005 - Testes | -1.44162 |
| | | | OBAT-12003 - Ovaries | -0.609531 |
| | | | OBAT-12003 - Testes | -1.19085 |
| GE16862 | *mei-9* | meiotic 9 (mei-9) | Thera2005 - Ovaries | -0.774595 |
| | | | Thera2005 - Testes | -1.60825 |
| GE19943 | *mus304* | mutagen-sensitive 304 (mus304) | Mainland Tai6 - Ovaries | -0.830693 |
| | | | Mainland Tai6 - Female soma | -4.03084 |
| GE19033 | *spel1* | spellchecker1 (spel1) | Thera6 - Ovaries | 0.0556472 |
| | | | Thera6 - Testes | -1.55863 |
| | | | Thera2005 - Ovaries | 0.308494 |
| | | | Thera2005 - Testes | -0.0460433 |
| | | | Island Cascade-1916 - Testes | -1.29778 |
| | | | Mainland Tai6 - Ovaries | -0.20007 |
| | | | Mainland Tai6 - Testes | -1.0472 |
| | | | Mainland Tai6 - Male soma | -1.45087 |
| | | | Mainland Tai6 - Female soma | -3.15695 |
| | | | OBAT-12003 - Ovaries | 0.352557 |
| | | | OBAT-12003 - Testes | -1.2141 |
| tacc | *tacc* | transforming acidic coiled-coil protein (tacc) | Thera2005 - Ovaries | 0.217986 |
| | | | Thera2005 - Testes | -0.846894 |
| | | | Thera2005 - Male soma | 1.01845 |
| GE18360 | *CG5181* | uncharacterized protein (CG5181) | Thera6 - Ovaries | 1.79065 |
| | | | Thera6 - Testes | 1.5854 |